\documentclass{article}
\usepackage[top=1in, bottom=1in, left=1.5in, right=1.5in]{geometry}

\usepackage{subfig}
\usepackage[bookmarks=false]{hyperref}
\usepackage{graphicx}
\usepackage{longtable}
\usepackage{algpseudocode}
\begin{document}

\title{An Event-Driven Approach for Studying Gene Block Evolution in Bacteria}
\author{David C Ream , Asma R Bankapur and Iddo Friedberg 
\footnote{to whom correspondence should be addressed:
i.friedberg@miamioh.edu}\\
Department of Microbiology \\
Miami University\\ Oxford, OH \\USA}



\maketitle

\begin{abstract}

\textbf{Motivation:}
Gene blocks are genes co-located on the chromosome. In many cases, genes
blocks are conserved between bacterial species, sometimes as operons, when genes are
co-transcribed. The conservation is rarely absolute: gene loss, gain, duplication, block
splitting, and block fusion are frequently observed. An open question in bacterial molecular evolution
is that of the formation and breakup of gene blocks, for which several models have been proposed. 
These models, however, are not generally applicable to all types of gene blocks, and consequently cannot
be used to broadly compare and study gene block evolution. To address this problem we 
introduce an event-based method for tracking gene block evolution in bacteria. 

\textbf{Results:}
We show here that the evolution of gene blocks in proteobacteria can be described by a small set
of events. Those include the insertion of genes into, or the splitting of genes out of a gene block, 
gene loss, and gene
duplication. We show how the event-based method of gene block evolution allows us to determine
the evolutionary rate, and to trace the ancestral states of their formation.  We conclude that
the event-based method can be used to help us understand the formation of these important
bacterial genomic structures.

\textbf{Availability:} The software is available under GPLv3 license on
\\
\url{http://github.com/reamdc1/gene\_block\_evolution.git}\\
Supplementary online material:\\
\url{http://iddo-friedberg.net/operon-evolution}

\textbf{Contact:} Iddo Friedberg {i.friedberg@miamioh.edu}
\end{abstract}

\section{Introduction}

In bacterial and archaeal genomes, gene blocks are sequences of genes
co-located on the chromosome. The evolutionary
conservation of gene blocks is strikingly apparent between many genomes.
It may also be that  conservations across numerous taxa indicate that at least some
conserved blocks are operons: a special case of gene blocks where the genes
are co-transcribed to polycistronic mRNA and are often associated with a
single function, such as a metabolic pathway or a protein complex. It is
estimated that 5-25\% of bacterial genes reside in
operons \cite{Wolf2001Genome}. Typically operons are under the control of one
or more regulator proteins, which facilitate co-regulated transcription.
From an evolutionary point of view, there are several questions that are asked
about operons and gene blocks. How did these units evolve?  What confers
fitness upon genes in an operon or block structure as opposed to not
being neighboring? Are certain operons more or less evolutionarily conserved in
bacteria?  What affects the conservation of the operon or gene block
structure in different taxa?

Several models exist to explain gene block evolution (for more extensive reviews see
\cite{Fondi2009Origin,Martin2009Recurring}. One of the first models proposed for biopathway evolution
is the Natal or Retrograde model which proposed that genes are arranged in blocks and operons  due to
tandem gene duplications derived from the depletion of metabolites in the
environment \cite{Horowitz1945Evolution}. However, this model does not explain many operons which encode
for proteins that are not homologous. Early \textit{co-adaptation models} (reviewed
in: \cite{Stahl1966Evolution}) applied to operons propose that neighboring genes into operons would lower
the chances of co-adapted genes being separated by random recombination.  However, orthologous
replacements of operon genes have been observed, suggesting that preservation of co-localization of
co-adapted alleles is not an exclusive reason for operons to form.  The \textit{coregulation model} is
derived from the original definition of an operon: that the neighboring operon genes is due to the
increased benefit of coregulation, providing an increased fitness for the population which has the
operon \cite{Price2005Operon}.  However, intermediate stages involving the cotranscription of
non-beneficial genes cannot explain an incremental increase in fitness. The \textit{selfish operon}
model \cite{Lawrence1996Selfish} proposes that the formation of gene blocks in bacteria is mediated by
transfer of DNA within and among taxa. The model proposes an increase in fitness for the constituent
genes because it enables the transfer of functionally coupled genes that would otherwise not increase
fitness if they were separate. Furthermore, the joining of genes into blocks and eventually operons is
beneficial for the horizontal gene transfer (HGT) of weakly selected, functionally coupled genes.  Thus, we
expect to see a certain percentage of ``genetic hitchhikers'': non-beneficial genes that are coupled to
beneficial genes in the operon. It seems that the selfish operon model does account for the structure of
some operons, but is not the only mechanism of operon construction. The main finding against the selfish
operon model's exclusivity is the much lower number of ``hitchhiking'', non-essential genes than
expected \cite{Pal2004Evidence}.  Price \emph{et al.} proposed that operon evolution is being driven by
selection on gene expression patterns, and they also found that although genes within operons are
usually closely spaced, genes in highly expressed operons may be widely spaced because of regulatory
fine-tuning by intervening sequences. This study was based on a comparative analysis of two genomes, but
included extensive expression data \cite{Price2006LifeCycle}.  Another model is that of the \emph{mosaic
operons} \cite{Omelchenko2003Evolution}. In this model, shuffling, disruption, and 
HGT play dominant parts in operon formation. Under this model, an operon is not a
steady-state evolutionary entity, but rather a dynamic entity which continuously acquires or loses genes
via HGT. In their paper, Omelchenko \emph{et al.} have shown that although some operons follow the
Selfish model, many do not, with HGT of individual genes into operons being quite common. Whole operon
transfer was identified in about 30\% of the operons studied. Another 20\% of the operons were
identified as mosaic operons.  However, the study does not attempt to further classify the different
types of mosaic operons, but rather provides an in-depth study of some of them. A study of the
\textit{his} operon by Fani \textit{et al.} has proposed a ``piecewise'' model to operon
evolution \cite{Fani2005Origin}. The piecewise model suggests that the construction of the \textit{his}
operon is a sequential series of events starting from a scattered set of constituent genes. Other models
include an adaptive life cycle of operons, in which they rarely evolve
optimally \cite{Price2006LifeCycle}.

Each of the operon evolution models present a mechanism and fits a biological rationale to the
observation that operons/gene blocks exist in extant taxa. However these models
do not readily allow us to quantify the changes between 
either operons/gene block or between different organisms. Moreover, more than one model 
can generally be applied to a chosen gene block and set of taxa.
Therefore, there is a need to create a universally applicable method
for charting gene block evolution. Having such a method on hand can help
determine the specific evolutionary trajectory of any given gene block.

\section{Approach}

Here we present a novel approach to investigate gene block evolution
which we call the \textit{event-driven method}. Our approach borrows from the
model describing the evolution of DNA and protein sequences. 
The accepted model for sequence evolution postulates two types of basic
events leading to changes: indels and mutations.  Indels and mutations are
assigned scores \cite{Dayhoff1978Atlas, Henikoff1992Amino} based on the frequency
of their occurrence over time. Given a pair of sequences, a typical hypothesis is posed 
as to whether they are homologs. The hypothesis is not rejected if we can show that
these two sequences are significantly similar. In
practical terms, the similarity between two sequences is ascertained if the
cumulative score of indels and mutation events differentiating the sequences is
below a certain threshold, determined by an appropriate null model, so that it
can be stated that the sequences are significantly similar. Note that the
smallest unit in which a change can happen is the nucleotide (DNA) or the
amino-acid (protein).

\begin{figure}[!tpb]
\centerline{\includegraphics[width=0.5\textwidth]{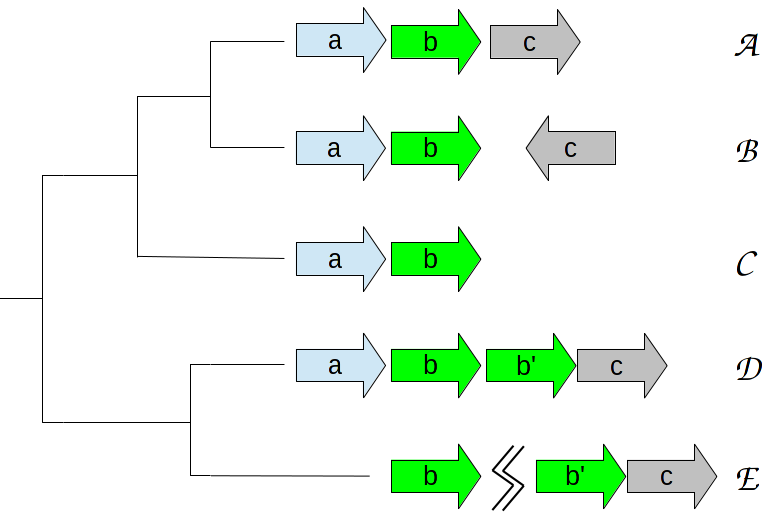}}

\caption{\footnotesize{The event-driven approach for operon evolution. 
Species $\mathcal{A-E}$ are arranged in a phylogenetic species tree. $\mathcal{A}$ is a source taxon,
with gene block $\mathcal{A}(a,b,c)$. 
In species B there is a strand reversal of the homolog $\mathcal{B}c$, which is treated as a split.
The orthoblock in species $\mathcal{C}$ has a deletion of gene $\mathcal{C}c$ when compared with
species $\mathcal{A}$ or $\mathcal{B}$. 
The orthoblock in $\mathcal{D}$ has a duplication of gene $\mathcal{B}b$ in relation to taxon
$\mathcal{A}$. The orthoblock in species $\mathcal{E}$ has a
split and a deletion of gene $\mathcal{E}a$.
}}

\label{fig:event_model} 
\end{figure} 

The event-driven method of gene block evolution we present here describes evolutionary events that occur between gene
blocks that are homologous between different bacterial species. The atomic unit of change is now the \textit{gene} as a
building-block of a gene block, rather than the nucleotide as the building block of a gene.  This procedure is best explained
by example. Suppose that genome $\mathcal{A}$ has neighboring genes $\mathcal{A}(a,b,c)$ in that order (In this annotation,
upper case letters are the taxon, lower case letters are the genes).  Genome $\mathcal{B}$ has homologs to those in
$\mathcal{A}$ $\mathcal{B}(a,b)$ are neighboring, but $\mathcal{B}c$ is located somewhere else in the chromosome, and
reversed. As for
genome $\mathcal{C}$, $c$ was deleted, so $\mathcal{C}(a,b)$ are neighboring. For the
scenario described, we can say that there was a gene split event between $\mathcal{A}$ and $\mathcal{B}$, and a gene
deletion event between $\mathcal{C}$ and any of the other genomes for the gene block $\mathcal{A}(a,b,c)$.  When the
phylogenetic tree is known, these events can be placed on the tree. See full example in
Figure~\ref{fig:event_model}.


If changes in gene blocks can be represented using a small set of
events, the number and type of these events can be used to describe the evolutionary
history of the gene block.  Here we report on a set of 38 operons from \textit{E.
coli} whose homologs we have examined across 33 taxa of proteobacteria. For these 38
operons and related orthologous gene blocks, we show an implementation of the event-driven approach to operon evolution. 


\section{Methods}

We define the following concepts: \textbf{reference taxon} is a taxon where operons have been
identified by experimental means.  Here we use \textit{E. coli} K-12 MG1655 as the reference
taxon. We chose \textit{E. coli} because it is expertly and comprehensively annotated in the
RegulonDB database \cite{Salgado2006RegulonDB}. \textbf{Neighboring genes}: two genes are
considered neighboring if they are 500 nucleotides or fewer apart, and on the same strand. A
\textbf{gene block} comprises no fewer than two open reading frames, or ORFs which are
neighboring. An \textbf{event} is a change in the gene block between any two species with
homologous gene blocks.  \textbf{Orthoblocks} (gene blocks that are orthologous) are defined as
follows: two organisms have orthoblocks when each organism must has at least two neighboring
genes that are homologous to genes in a gene block in the reference taxon's genome. Genes are
considered homologous if their pairwise BLAST e-value is $10^{-10}$ or less. Relying on a
strict BLAST threshold may exclude homologous proteins whose
sequence similarity is not high (false negatives). However, this strategy will rarely include
proteins with a different function (false positives). This rigorous threshold was chosen with
the primary goal of minimizing false positives when inferring function by similarity. 

\subsection{Evolutionary events}

Next we define events that we use to examine changes in gene block structure between different
bacterial taxa relative to \textit{E. coli}.  We chose a set of target taxa with known
phylogenetic relationships. The genomes of the target taxa were searched for homologous blocks
to the operons found in \textit{E. coli}. The operons we chose were selected based on the following
criteria: 1. all the genes were protein coding;  2. for all blocks chosen, the co-transcription
was experimentally determined in \textit{E. coli}; 3. each operon comprised at least five genes;
4. each operon has  orthoblocks in at least nine other genomes.
Using these filtering criteria, and  RegulonDB's annotation of \textit{E. coli} as our reference
taxon, we compiled 38 operons for this study. See Supplementary Material,
\href{http://iddo-friedberg.net/operon-evolution}{Table~S1} for a full list of operons.

We define the following pairwise events between orthoblocks from different taxons:

\begin{enumerate} 

    \item \textbf{Splits} If two genes in one taxon are neighboring and their homologs in the
    other taxon are not, then that is defined as a single \textit{split event}. The distance is
    the minimal number of split events identified between the compared genomes.

    \item \textbf{Deletions} A gene exists in the operon in the one taxon, but its homolog
    cannot be found in an orthoblock in another taxon. Note that the definition of homolog,
    e-value $10^{-10}$ is strict, and may result in false negatives. The \textit{deletion
    distance} is the number of deletion events identified between the compared target genomes.

    \item \textbf{Duplications} A duplication event is defined as having gene
    $j$ in a gene block in the source genome, and a homologous genes
    $(j',j'')$ in the homologous block in the target genome. The
    \textit{duplication distance} is the number of duplication events counted
    between the source and target genomes. The duplication has to occur in a
    gene block to be tallied.


\end{enumerate}

Other events were examined too: rearrangement of genes, genes moving to another strand, fusion and
fission of open reading frames. These event types correlated strongly with one or more of the three
event types listed above, and were therefore discarded. Fusion/fission of open reading frames were rare
rare in our data set, so this event type was discarded as well.

The event-driven method does not account for horizontal gene transfer, which is suggested as a common
mechanism for transferring neighboring genes \cite{Lawrence1996Selfish}. However, We have not yet
incorporated HGT into our model. We have tried using
AlienHunter \cite{Langille2008Evaluation}, and the IslandViewer \cite{Langille2009IslandViewer} suite to
detect HGT events in our data.  However, given that the taxa we are analyzing are closely related,
these software were unable to detect HGT events. 

\subsection{Different Conservation Rates for Gene Blocks in Proteobacteria}

\textbf{Determining orthology:} To trace the events that affect genes in gene blocks, it is necessary to determine which genes
are orthologous between any two taxa when more than two possible homolog
pairings may exist. The problem may be stated as follows: given a gene $g$ in genome $A$, and a set of
homologs to $g$ in genome $B$, $H^B_g=\{g_1, g_2,...,g_n\}$, which of the genes in $H^B_g$ is
the ortholog to $g$?  The Best Reciprocal Hits (BRH) method is commonly used to find orthologs,
however, BRH assumes that ortholog $g_i$ is necessarily that which is most similar to $g$,
discounting the possibility of different evolutionary rates of paralogs. We therefore take a
different approach in determining ortholog identity for genes in homology blocks. When
selecting a single ortholog among all possible homologs in $H_g$, we use synteny and
sequence similarity to determine which of the genes in an examined genome is the correct ortholog.
To do so we use the following three criteria:

\begin{enumerate}

\item \textbf{Prioritizing by gene blocks} We prioritize orthologs that are in gene blocks over
orthologs that are isolated in the genome, and we look for the minimal number of such blocks
that contain a representative of every ortholog that we recover.  Example: the operon in
\textit{E. coli} had gene block $(abcdef)$. The target genome has the following orthologs
grouped in its genome: $(abcd)$, $(abc)$. In this case, we will choose as orthologs the genes
populating $(abcd)$. 

\item \textbf{Recovering maximum number of genes} We consider the number of genes found.
Example:  the reference taxon operon had the gene block $(abcdef)$. The studied genome has the
following blocks $(abcd)$ and $((abc),(de))$. We would choose $((abc),(de))$, two blocks,even
though $(abcde)$ is one block, because in the latter case we recover more homologs.

\item \textbf{Minimizing duplications} If in the target genome we have a choice between ortholog
groups $((abc),(de))$ or $((abcd), (de))$ we choose the first because it has the minimal number
of gene duplications.
\end{enumerate}

We now define a \textit{target homolog} as a gene in the target genome that is a homolog to a gene
in a gene block in the reference genome \textit{E. coli}. and a \textit{target homolog block} as
one or more target homologs, spaced $\le 500$ bp.

\begin{algorithmic}[1]
\For{geneBlock in ReferenceGenome}
	\For{genes in  geneBlock}
		\State Find all homologs in the target genome with BLAST e-value $\le 10^{-10}$
	\EndFor
	\State Find all homologs in the target genome that are neighboring ($\le500$bp)
	\State Use a greedy algorithm to recover the maximum number of target homologs prioritizing by gene blocks while
    minimizing gene duplications and maximizing number of genes recovered.
\EndFor
\end{algorithmic}

\textbf{Event-based distances}. Once orthologs are chosen, we are able to define the event-based distance between any
two gene blocks with respect to split events, duplication events and deletion events. The distance between any two
homologous gene blocks found in target organisms is defined as follows: 

\begin{enumerate}

    \item \textit{Split distance} ($d_s$) is the absolute difference in the number of relevant gene
    blocks between the two taxa.  Example: for the reference gene block with genes $(abcdefg)$
    Genome $A$ has blocks $((abc), (defg))$ and genome $B$ has $((abc), (de),(fg))$. Therefore,
    $d_s(A,B) = |2-3| = 1$. 
    
    \item \textit{Duplication distance} ($d_u$) is the pairwise count of duplications between two
    orthoblocks. 
    Example: we have a reference gene block
    $(abcde)$. Now, for genomes A and B the orthoblocks are $A=((abd))$ $B=((abbcc))$.
    Gene $b$ causes a duplication distance $d_u(A,B)$ of 1. Gene $c$ generates a distance of one
    deletion (see below) and one duplication. This is because the most parsimonious explanation is
    that the most recent ancestor for $A$ and $B$  may have had one copy of $c$, thus generating a
    duplication in one lineage, and a deletion in another. Since gene $d$ exists only in the
    reference genome, it has no bearing on the event-based distance between the homologous gene
    blocks $A$ and $B$.  
    
    \item \textit{Deletion distance} $d_d$ is the difference in number of orthologs that are in the homologous
    gene blocks of the genome of one organism, or the other, but not in both. 

\end{enumerate}

\textbf{Gene block frequency matrices} The rules outlined above allow us to determine the
pairwise distance, for a given event and gene block, between any two genomes in our corpus. To
visualize the frequency of an event in a block, we created matrices whose axes are the
examined species, and whose cell is the normalized value of the pairwise distance for any given event. 
For an event $v$ being one
of insertion, deletion or duplication, for any two taxa $i$ and $j$ with homology blocks, the
value for the normalized distance matrix entry $M_{ij}$ is:

\[
M_{ij} = \frac{d_v(i,j) - \bar{x}{_{d_v}}}{\sigma_{d_v}}
\]

Where $\bar{x}{_{d_v}}$ is the mean value of the distance for event $v$ calculated over all
pairs of taxa $n_p$ sharing that event:

\[
\bar{x}{_{d_v}} = \frac{1}{n_p} \sum\limits_{i<j}{d_v(i,j)}
\]

and $\sigma_{d_v}$ is the standard deviation.

\textbf{Choice of proteobacteria species, and phylogenetic tree construction}. We chose our species as
in \cite{Fani2005Origin}, removing a few species that we deemed to be too evolutionarily close. The phylogenetic trees
shown in the study were constructed from multiple sequence alignments of the \textit{rpoD} gene, using ClustalX
2.1 \cite{Larkin2007Clustal}, followed by neighbor-joining. Table~S3 lists the species used.


\section{Results}

\begin{figure*}[htbp]
    \subfloat[rplKAJL-rpoBC: deletions, duplications,
    splits]{\includegraphics[width=1.0\textwidth]{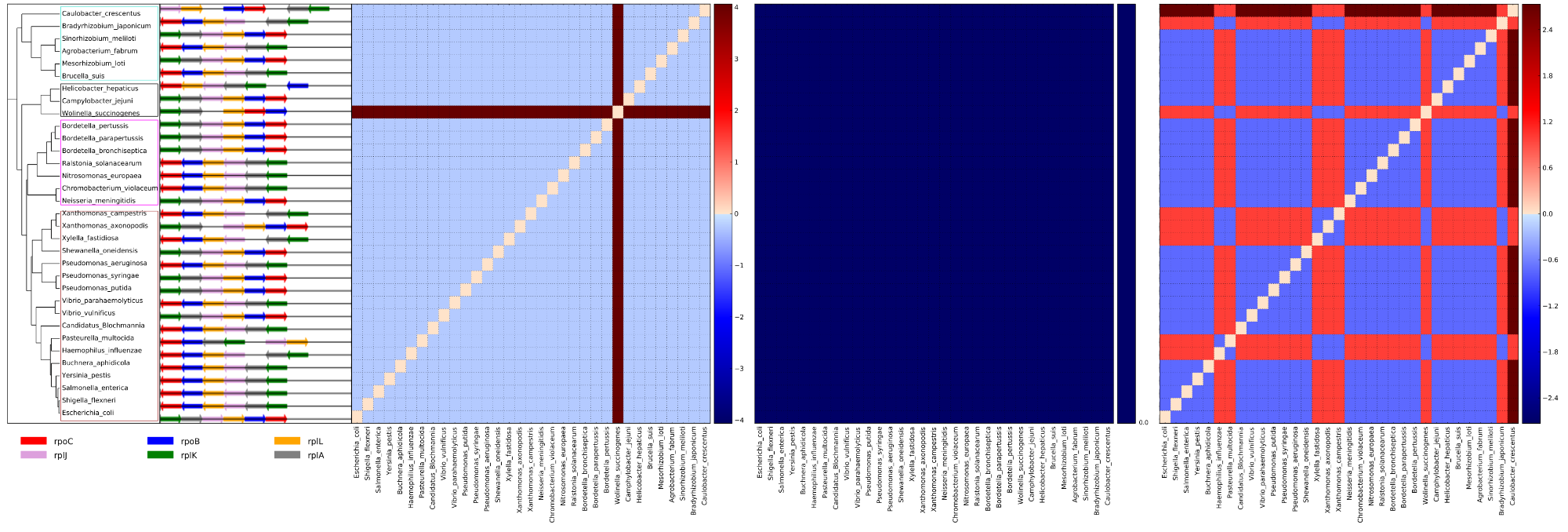}}
    \\
    \subfloat[atpIABCDEFGH: deletions, duplications,
    splits]{\includegraphics[width=1.0\textwidth]{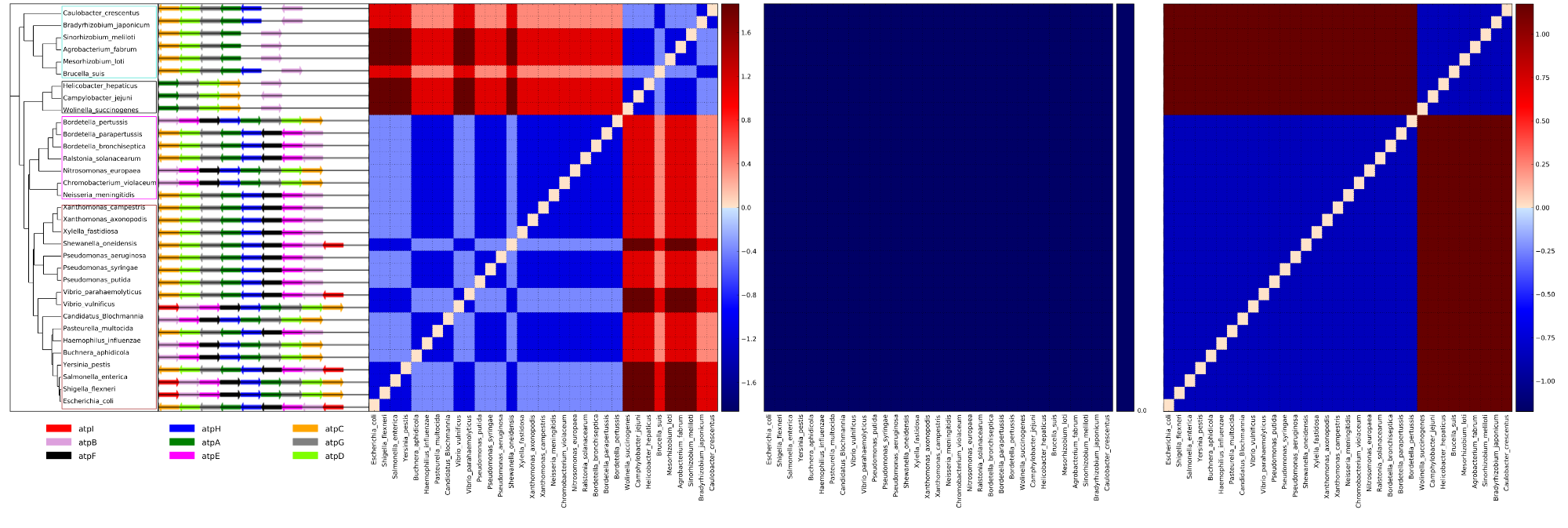}}
\caption{\footnotesize{Highly conserved orthoblocks. 
The color matrices each show degree of relative conservation of the event between any
two species. Left to right: Deletions, duplications, splits. Blue to red scale is high to low conservation \textit{z}-score as 
described in
Methods. The boxes outline (top to bottom): $\alpha$-, $\epsilon$-, $\beta$-, and $\gamma$-proteobacteria.
\textbf{a}: rplKAJL-rpoBC has only a single gene deletion in \textit{Wollinela}, no gene duplications, and a few splits (red
squares, rightmost panel) including genes that moved to another strand. \textbf{b}: the atp orthoblock shows deletions of 
\textit{atpI}
and a false deletion of \textit{atpE} due to low similarity to \textit{E. coli atpE} in $\epsilon$ and $\alpha$ proteobacteria (left
matrix). No gene duplications are exhibited (middle panel). Splits are due to strand reversal of component genes (right
panel. High resolution figure available at: \url{http://iddo-friedberg.net/operon-evolution/}}}
\label{fig:blocks1} 
\end{figure*} 

\begin{figure*}[htbp]
    \subfloat[paaABCDEFGHIJK: deletions, duplications,
    splits]{\includegraphics[width=1.0\textwidth]{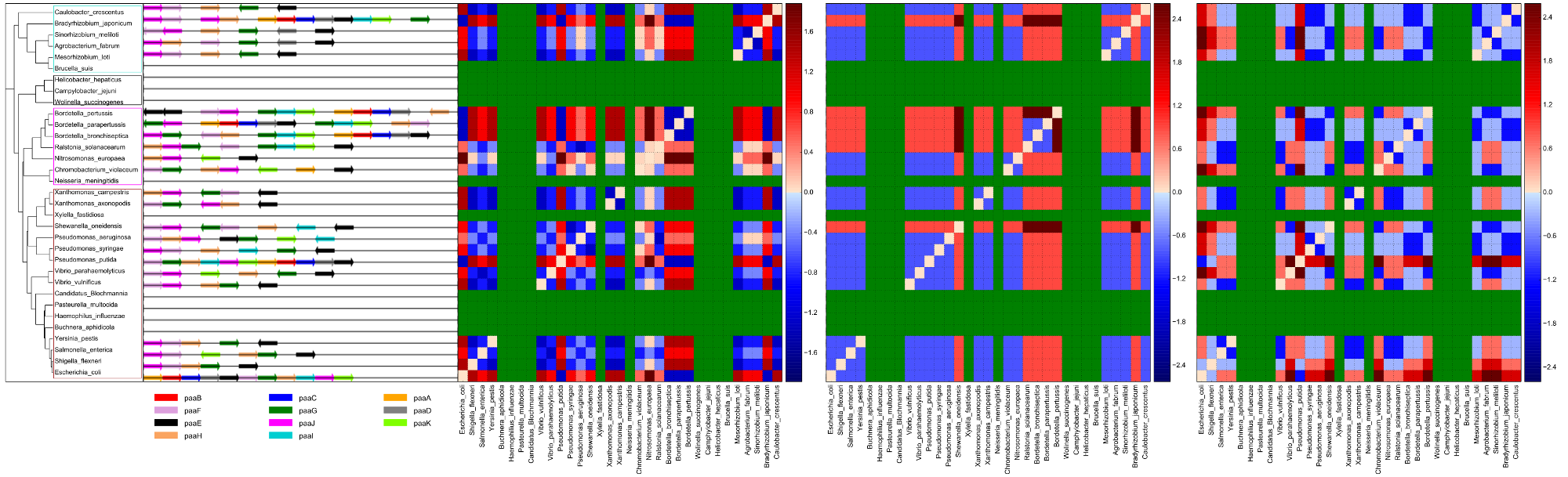}}
    \\
    \subfloat[hyfABCDEFGHIJR-focB: deletions, duplications,
    splits]{\includegraphics[width=1.0\textwidth]{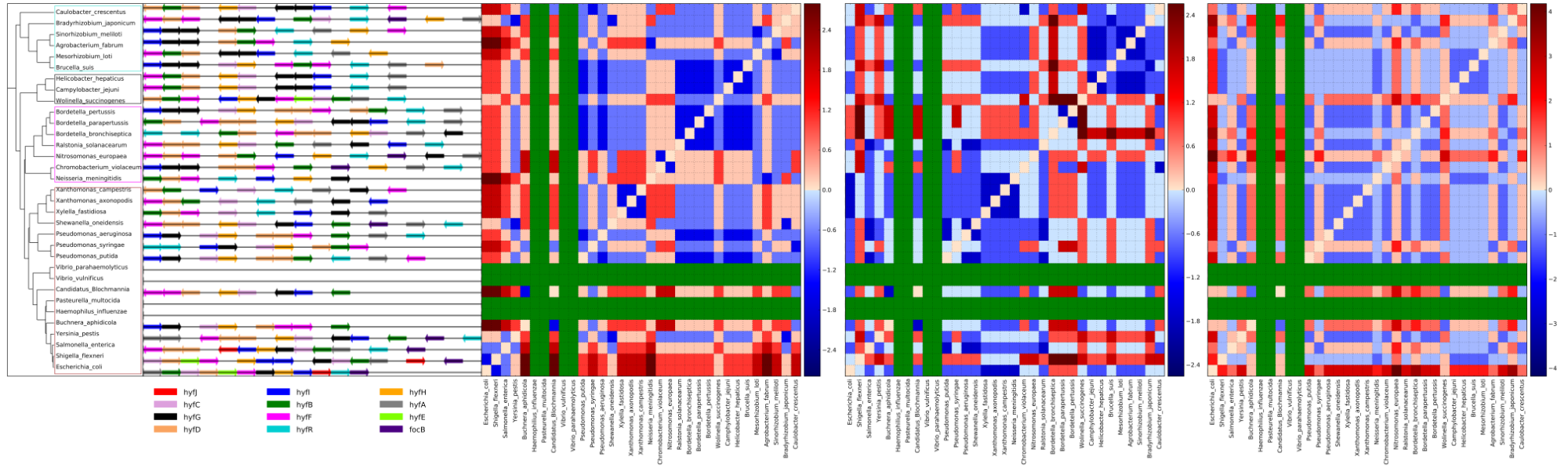}}
\caption{\footnotesize{Less conserved orthoblocks. Green squares are genomes in which component genes were not found using
BLAST. \textbf{a}: phenylacetate degradation orthoblock. \textbf{b}: the hyf operon encoding the fourth hydrogenase in \textit{E. coli}
is not expressed under known conditions. See text for details on both operons. 
High resolution figure available at: \url{http://iddo-friedberg.net/operon-evolution/}}}
\label{fig:blocks2} 
\end{figure*} 
The results of this study show an interesting variety in gene block evolution.  First, we show
the \textit{gene block tree} diagrams in a phylogenetic tree. These diagrams show the gene
blocks as we find them in the different species we examine. The lack of a gene in the gene block
phylogenetic tree does not mean the homolog does not exist in that taxon, but rather that it is
no longer detectable by BLAST at the threshold of $10^{-10}$. Further, if a gene block is missing
from the phylogenetic tree, it means that there are no two genes in the gene block that are 
neighboring in the genome within a distance of $\le 500$bp.

To visualize the frequency of events, we generate gene block event frequency matrices as
described in Methods. The value of each matrix is a \textit{z}-score. Figure~\ref{fig:blocks1} shows two conserved 
gene blocks, and Figure~\ref{fig:blocks2} shows two non-conserved gene blocks.


\subsection{Conservation of Gene Blocks and Relationship to Function}

\begin{figure*}[!tbp]
    \includegraphics[width=0.8\textwidth]{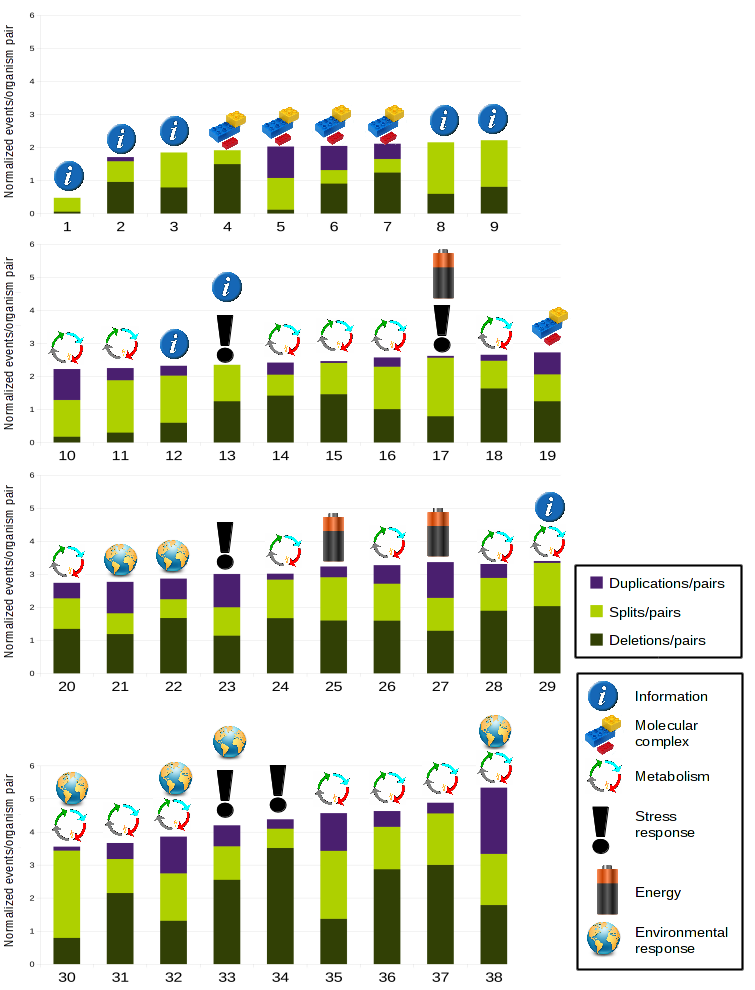}
\caption{\footnotesize{Relative conservation of operons and their primary biological functions. Each bar shows the cumulative number of
events per genome pair per orthoblock. The orthoblocks are numbered as in \href{http://iddo-friedberg.net/operon-evolution}{Table~S1}. Conserved orthoblocks have
shorter bars, operons are ordered top-to-bottom left-to-right from most conserved to least conserved.
}}
\label{fig:function} 
\end{figure*}

The event-driven method enables us to examine the relative conservation of gene blocks in proteobacteria. Figure
\ref{fig:function} the gene blocks are arranged in descending order of conservation. 
The most conserved
block is the operon rplKAJL-rpoBC, a highly conserved transcription unit of ribosomal proteins (\textit{rplK, rplK,
rplA} and \textit{rplL}) and two RNA polymerase subunits (\textit{rpoB} and \textit{rpoC}) \cite{Steward1991In}, 
Figure~\ref{fig:blocks1}(a).

No gene duplications or deletions
were detected.  Our program does erroneously call a deletion of the rplJ in \textit{Wolinella}, but this
is an error due to the stringent e-value cutoff and having a single exemplar of the \textit{rplJ} gene as a query. The
splits we detect are mostly between the \textit{rplKA} and \textit{rplJL-rpoBC} transcription units. The genes in this operon
have multiple promotor and attenuator sites \cite{Ralling1984Relative}, and have been shown to be governed by a
complex set of signals \cite{Steward1991In}. It appears that a complete tetracistronic
product is transcribed from rplKAJL with less abundant bicistronic products of \textit{rplK-rplA} and 
\textit{rplJ-rplL}
\cite{Downing1987Transcription},
which may explain the strong conservation of these four genes in this gene block.



Another well-conserved operon is the atp operon, which codes for the genes for the $H^+$-ATPase complex.  ATP
synthase is responsible for generating ATP using the proton motive force (PMF) across the cell
membrane \cite{FERNANDEZMORAN1964MACROMOLECULAR}. 
We examined the operon coding for ATP synthase, atpIBEFHAGDC. While highly conserved, this operon does exhibit
gene deletions in some taxa.  Most notably the gene atpI, a nonessential gene that codes for a helper protein that assists the assembly
of the ATP-synthase complex's rotor.  We readily recover this gene in orthoblocks in organisms that are closely
related to \textit{E.coli}. 
We did not observe any duplications in our dataset, 
The atpI deletion was a true deletion,
which makes sense functionally as the \textit{atpI} gene codes for the AtpI protein which is a nonessential component of the
$H^+$-ATPase complex. 
The other components supposedly deleted, \textit{afpF, atpE} in $\epsilon$-proteobacteria 
and $\alpha$-proteobacteria are
highly dissimilar to the equivalent \textit{E. coli} genes, and are therefore not identifiable as homologs (e-value
using BLAST $> 0.01$, data not shown). See Figure~\ref{fig:blocks1}(b).

At the other edge of the conservation spectrum, we examined the hyf operon. As the genome diagram shows, the gene
block of 12 genes is not conserved as a single block in any of the proteobacteria in our data
set \cite{Andrews199712cistron}. The hyf proton-translocating formate hydrogenlyase block of 12 genes appears to be an
operon only in \textit{E. coli}, although many of the genes appear in separate blocks in other bacteria. The hyf
operon in \textit{E. coli} is probably silent, at least under the environmental conditions examined, and has only
been expressed under artificial conditions \cite{Self2004Expression}. Not being able to express it in \textit{E.
coli} under native conditions suggests it may be redundant, as does its lack of conservation in the species
examined. See Figure~\ref{fig:blocks2}(a).


The paa operon in \textit{E. coli} encodes for a multicomponent oxygenase/reductase subunit for the aerobic degradation of
phenylacetic acid. The \textit{E. coli} operon comprises 11 genes. The distribution of the genes of this operon in 102
bacterial genomes was studied in detail in \cite{Martin2009Recurring}. The authors' conclusions from this study was that
\textit{de novo} clustering of some of this orthoblock's genes occur repeatedly, due to weak selective pressure. The
proximity of genes sets up opportunities for co-transcription. Specifically, this study has shown that genes \textit{paaA,
B, C} and \textit{paaD}, when they are found, always co-occur in an operon. This makes sense, as those genes form a stable
molecular complex with those genes coding for essential subunits for the degradation of
phenylacetate \cite{Grishin2011Structural}. Both our study's and Martin \textit{et al}'s found the full gene block  only in
\textit{E. coli} and \textit{Pseudomonas putida}.  Another gene-block co-occurrence we find in our study is that of paaF
and paaG, in 12 out of 23 species in which any components of the paa orthoblock occur. The products of these two genes form
the FaaGH complex, another stable complex which catalyzes consecutive steps in the phenylacetate degradation pathway, and
it was hypothesized that the proximity of the two proteins in a complex provides a fitness
advantage \cite{Grishin2012ProteinProtein}. 
Another use of the event-driven method is the reconstruction of ancestral gene blocks along the
evolutionary tree. \href{http://iddo-friedberg.net/operon-evolution}{Figure~S1} shows such a reconstruction for
paa in the $\gamma$-proteobacteria
species used in our study. We manually examined the possible events needed to transition between the tree's
nodes, and along each branch minimized the number of events leading to the extant orthoblocks. One interesting
outcome of this analysis, is that paa orthoblock appears to be the result of HGT events in \textit{E. coli} and
in \textit{P. putida}, where an entire gene block exists in both species, but not in the closely related ones.

To determine if there is a relationship between gene block conservation and the function of
the gene blocks, we assigned each operon keywords based on its function. The categories we
used were Metabolism, Information, Molecular Complex, Stress Response, Energy, and
Environmental Response. The keywords were assigned based on reading the literature
relevant for each operon, and the information provided in EcoCyc \cite{Keseler2013EcoCyc}. As can be
seen in Figure~\ref{fig:function}, from the gene blocks we studied, gene blocks whose
function is information or protein complex tend to be more conserved, and those dealing
with stress response are less conserved. Gene blocks whose primary function was identified
as metabolism were found at all levels of conservation. We conclude that there may be a
relationship between gene block conservation and function, and that gene blocks having to
do with information or molecular complexes are more conserved than those dealing with
stress response and/or environmental response.

\section{Discussion and Conclusions}

We introduce a method to examine gene block and operon evolution in bacteria. This method, coupled with the
visualization we present, enables the interrogation of the evolution of gene blocks in a bacterial clade. The
event-driven approach we use allows for the quantification of evolutionary conservation of any gene block.
Most importantly, the event-driven method does not attempt to present a predictive
evolutionary model such as the models reviewed in the introduction. Rather,
it allows for these models to be examined in specific gene blocks with specific taxa. The event-driven method is
agnostic to any model predicting how an operon or gene block evolved.


To determine the conservation of gene blocks, a choice needs to be made for the proper orthologs between genomes.
Identifying orthologs is a challenging problem that has been studied extensively. Several methods have been developed to
do so including use of inter-species clusters \cite{Tatusov1997Genomic, Powell2014EggNOG, Remm2001Automatic}, reciprocal
best hits \cite{Tatusov1997Genomic}, or phylogenetic methods \cite{Fulton2006Improving}. Here we used genomic context and
strict similarity criteria to choose which genes are orthologous, and are likely to have the same function. The ortholog
choice method we use here assumes that evidence for orthology is strengthened by genomic context. This assumption has been
shown to be useful for a more precise assignment of orthologs \cite{Jun2009Identification,Ward2014Quickly} and has been
implemented in computational resources \cite{Szklarczyk2014STRING, Aziz2008RAST,Overbeek2014SEED} to resolve ortholog
ambiguities. 

We choose gene blocks on the basis that, at least in one species (\textit{E. coli}), the genes are co-transcribed.
The initial motivation for this study was the observation that orthoblocks have been
shown to be useful in inferring common function, even when co-transcription has not been determined
\cite{Overbeek1999Use,Enault2005Phydbac}. We have shown that tracking gene block evolution in bacteria through the
tallying of simple events provides an obejective, quantifiable method for understanding their evolutionary
conservation, relative to a reference species. We have shown examples of two highly conserved orthoblocks (atp and ydc)
and two less-conserved orthoblocks (paa and hyf). We have also related overall conservation to the type of orthoblock
function, although a larger survey of orthoblocks is needed to obtain a more reliable picture.

\paragraph{Funding:} This work is supported, in part, by the 
US National Science Foundation under Grant Number
ABI-1146960. Any opinions, findings, and conclusions or recommendations 
expressed in this material are those of the authors and do not necessarily
reflect the views of the National Science Foundation.

\bibliographystyle{plain}
\bibliography{operon_evolution_caps}

\begin{thebibliography}{10}

\bibitem{Andrews199712cistron}
Simon~C. Andrews, Ben~C. Berks, Joseph McClay, Andrew Ambler, Michael~A. Quail,
  Paul Golby, and John~R. Guest.
\newblock {A 12-cistron Escherichia coli operon (hyf) encoding a putative
  proton-translocating formate hydrogenlyase system}.
\newblock {\em Microbiology}, 143(11):3633--3647, November 1997.

\bibitem{Aziz2008RAST}
Ramy~K. Aziz, Daniela Bartels, Aaron~A. Best, Matthew DeJongh, Terrence Disz,
  Robert~A. Edwards, Kevin Formsma, Svetlana Gerdes, Elizabeth~M. Glass,
  Michael Kubal, Folker Meyer, Gary~J. Olsen, Robert Olson, Andrei~L. Osterman,
  Ross~A. Overbeek, Leslie~K. McNeil, Daniel Paarmann, Tobias Paczian, Bruce
  Parrello, Gordon~D. Pusch, Claudia Reich, Rick Stevens, Olga Vassieva,
  Veronika Vonstein, Andreas Wilke, and Olga Zagnitko.
\newblock {The RAST Server: rapid annotations using subsystems technology.}
\newblock {\em BMC genomics}, 9(1):75+, February 2008.

\bibitem{Dayhoff1978Atlas}
M.O. Dayhoff.
\newblock {\em Atlas of Protein Sequence and Structure: supplement 3 1978}.
\newblock Number v. 5. National Biomedical Research Foundation, 1978.

\bibitem{Downing1987Transcription}
Willa~L. Downing and Patrick~P. Dennis.
\newblock {Transcription products from the rplKAJL-rpoBC gene cluster}.
\newblock {\em Journal of Molecular Biology}, 194(4):609--620, April 1987.

\bibitem{Enault2005Phydbac}
Francois Enault, Karsten Suhre, and Jean~M. Claverie.
\newblock {Phydbac "Gene Function Predictor" : a gene annotation tool based on
  genomic context analysis}.
\newblock {\em BMC Bioinformatics}, 6(1):247+, 2005.

\bibitem{Fani2005Origin}
Renato Fani, Matteo Brilli, and Pietro Li\`{o}.
\newblock {The origin and evolution of operons: the piecewise building of the
  proteobacterial histidine operon.}
\newblock {\em Journal of molecular evolution}, 60(3):378--390, March 2005.

\bibitem{FERNANDEZMORAN1964MACROMOLECULAR}
H.~Fernandez~Moran, T.~Oda, P.~V. Blair, and D.~E. Green.
\newblock {A macromolecular repeating unit of mitochondrial structure and
  function. Correlated electron microscopic and biochemical studies of isolated
  mitochondria and submitochondrial particles of beef heart muscle.}
\newblock {\em The Journal of cell biology}, 22:63--100, July 1964.

\bibitem{Fondi2009Origin}
Marco Fondi, Giovanni Emiliani, and Renato Fani.
\newblock {Origin and evolution of operons and metabolic pathways}.
\newblock {\em Research in Microbiology}, 160(7):502--512, September 2009.

\bibitem{Fulton2006Improving}
Debra Fulton, Yvonne Li, Matthew Laird, Benjamin Horsman, Fiona Roche, and
  Fiona Brinkman.
\newblock {Improving the specificity of high-throughput ortholog prediction}.
\newblock {\em BMC Bioinformatics}, 7(1):270+, May 2006.

\bibitem{Grishin2011Structural}
Andrey~M. Grishin, Eunice Ajamian, Limei Tao, Linhua Zhang, Robert Menard, and
  Miroslaw Cygler.
\newblock {Structural and functional studies of the Escherichia coli
  phenylacetyl-CoA monooxygenase complex.}
\newblock {\em The Journal of biological chemistry}, 286(12):10735--10743,
  March 2011.

\bibitem{Grishin2012ProteinProtein}
Andrey~M. Grishin, Eunice Ajamian, Linhua Zhang, Isabelle Rouiller, Mihnea
  Bostina, and Miroslaw Cygler.
\newblock {Protein-Protein Interactions in the Î²-Oxidation Part of the
  Phenylacetate Utilization Pathway}.
\newblock {\em Journal of Biological Chemistry}, 287(45):37986--37996, November
  2012.

\bibitem{Henikoff1992Amino}
S.~Henikoff and J.~G. Henikoff.
\newblock {Amino acid substitution matrices from protein blocks}.
\newblock {\em Proceedings of the National Academy of Sciences},
  89(22):10915--10919, November 1992.

\bibitem{Horowitz1945Evolution}
N.~H. Horowitz.
\newblock {On the Evolution of Biochemical Syntheses}.
\newblock {\em Proceedings of the National Academy of Sciences},
  31(6):153--157, June 1945.

\bibitem{Jun2009Identification}
Jin Jun, Ion Mandoiu, and Craig Nelson.
\newblock {Identification of mammalian orthologs using local synteny}.
\newblock {\em BMC Genomics}, 10(1):630+, 2009.

\bibitem{Keseler2013EcoCyc}
Ingrid~M. Keseler, Amanda Mackie, Martin Peralta-Gil, Alberto Santos-Zavaleta,
  Socorro Gama-Castro, C\'{e}sar Bonavides-Mart\'{\i}nez, Carol Fulcher,
  Araceli~M. Huerta, Anamika Kothari, Markus Krummenacker, Mario Latendresse,
  Luis Mu\~{n}iz Rascado, Quang Ong, Suzanne Paley, Imke Schr\"{o}der,
  Alexander~G. Shearer, Pallavi Subhraveti, Mike Travers, Deepika Weerasinghe,
  Verena Weiss, Julio Collado-Vides, Robert~P. Gunsalus, Ian Paulsen, and
  Peter~D. Karp.
\newblock {EcoCyc: fusing model organism databases with systems biology}.
\newblock {\em Nucleic Acids Research}, 41(D1):D605--D612, January 2013.

\bibitem{Langille2008Evaluation}
Morgan Langille, William Hsiao, and Fiona Brinkman.
\newblock {Evaluation of genomic island predictors using a comparative genomics
  approach}.
\newblock {\em BMC Bioinformatics}, 9(1):329+, August 2008.

\bibitem{Langille2009IslandViewer}
Morgan G.~I. Langille and Fiona S.~L. Brinkman.
\newblock {IslandViewer: an integrated interface for computational
  identification and visualization of genomic islands}.
\newblock {\em Bioinformatics}, 25(5):664--665, March 2009.

\bibitem{Larkin2007Clustal}
M.~A. Larkin, G.~Blackshields, N.~P. Brown, R.~Chenna, P.~A. McGettigan,
  H.~McWilliam, F.~Valentin, I.~M. Wallace, A.~Wilm, R.~Lopez, J.~D. Thompson,
  T.~J. Gibson, and D.~G. Higgins.
\newblock {Clustal W and Clustal X version 2.0}.
\newblock {\em Bioinformatics}, 23(21):2947--2948, November 2007.

\bibitem{Lawrence1996Selfish}
J.~G. Lawrence and J.~R. Roth.
\newblock {Selfish operons: horizontal transfer may drive the evolution of gene
  clusters.}
\newblock {\em Genetics}, 143(4):1843--1860, August 1996.

\bibitem{Martin2009Recurring}
Fergal Martin and James McInerney.
\newblock {Recurring cluster and operon assembly for Phenylacetate degradation
  genes}.
\newblock {\em BMC Evolutionary Biology}, 9(1):36+, February 2009.

\bibitem{Omelchenko2003Evolution}
Marina Omelchenko, Kira Makarova, Yuri Wolf, Igor Rogozin, and Eugene Koonin.
\newblock {Evolution of mosaic operons by horizontal gene transfer and gene
  displacement in situ}.
\newblock {\em Genome Biology}, 4(9):R55+, 2003.

\bibitem{Overbeek1999Use}
R.~Overbeek, M.~Fonstein, M.~D'Souza, G.~D. Pusch, and N.~Maltsev.
\newblock {The use of gene clusters to infer functional coupling.}
\newblock {\em Proceedings of the National Academy of Sciences of the United
  States of America}, 96(6):2896--2901, March 1999.

\bibitem{Overbeek2014SEED}
Ross Overbeek, Robert Olson, Gordon~D. Pusch, Gary~J. Olsen, James~J. Davis,
  Terry Disz, Robert~A. Edwards, Svetlana Gerdes, Bruce Parrello, Maulik
  Shukla, Veronika Vonstein, Alice~R. Wattam, Fangfang Xia, and Rick Stevens.
\newblock {The SEED and the Rapid Annotation of microbial genomes using
  Subsystems Technology (RAST).}
\newblock {\em Nucleic acids research}, 42(Database issue):D206--D214, January
  2014.

\bibitem{Pal2004Evidence}
Csaba P\'{a}l and Laurence~D. Hurst.
\newblock {Evidence against the selfish operon theory}.
\newblock {\em Trends in Genetics}, 20(6):232--234, June 2004.

\bibitem{Powell2014EggNOG}
Sean Powell, Kristoffer Forslund, Damian Szklarczyk, Kalliopi Trachana,
  Alexander Roth, Jaime Huerta-Cepas, Toni Gabald\'{o}n, Thomas Rattei, Chris
  Creevey, Michael Kuhn, Lars~J. Jensen, Christian von Mering, and Peer Bork.
\newblock {eggNOG v4.0: nested orthology inference across 3686 organisms}.
\newblock {\em Nucleic Acids Research}, 42(D1):D231--D239, January 2014.

\bibitem{Price2006LifeCycle}
Morgan~N. Price, Adam~P. Arkin, and Eric~J. Alm.
\newblock {The Life-Cycle of Operons}.
\newblock {\em PLoS Genet}, 2(6):e96+, June 2006.

\bibitem{Price2005Operon}
Morgan~N. Price, Katherine~H. Huang, Adam~P. Arkin, and Eric~J. Alm.
\newblock {Operon formation is driven by co-regulation and not by horizontal
  gene transfer}.
\newblock {\em Genome Research}, 15(6):809--819, June 2005.

\bibitem{Ralling1984Relative}
G.~Ralling and T.~Linn.
\newblock {Relative activities of the transcriptional regulatory sites in the
  rplKAJLrpoBC gene cluster of Escherichia coli.}
\newblock {\em Journal of bacteriology}, 158(1):279--285, April 1984.

\bibitem{Remm2001Automatic}
Maido Remm, Christian E.~V. Storm, and Erik L.~L. Sonnhammer.
\newblock {Automatic clustering of orthologs and in-paralogs from pairwise
  species comparisons}.
\newblock {\em Journal of Molecular Biology}, 314(5):1041--1052, December 2001.

\bibitem{Salgado2006RegulonDB}
Heladia Salgado, Socorro Gama-Castro, Mart\'{\i}n Peralta-Gil, Edgar
  D\'{\i}az-Peredo, Fabiola S\'{a}nchez-Solano, Alberto Santos-Zavaleta, Irma
  Mart\'{\i}nez-Flores, Ver\'{o}nica Jim\'{e}nez-Jacinto, C\'{e}sar
  Bonavides-Mart\'{\i}nez, Juan Segura-Salazar, Agustino Mart\'{\i}nez-Antonio,
  and Julio Collado-Vides.
\newblock {RegulonDB (version 5.0): Escherichia coli K-12 transcriptional
  regulatory network, operon organization, and growth conditions.}
\newblock {\em Nucleic acids research}, 34(Database issue), January 2006.

\bibitem{Self2004Expression}
William~T. Self, Adnan Hasona, and K.~T. Shanmugam.
\newblock {Expression and regulation of a silent operon, hyf, coding for
  hydrogenase 4 isoenzyme in Escherichia coli.}
\newblock {\em Journal of bacteriology}, 186(2):580--587, January 2004.

\bibitem{Stahl1966Evolution}
F.~W. Stahl and N.~E. Murray.
\newblock {The evolution of gene clusters and genetic circularity in
  microorganisms.}
\newblock {\em Genetics}, 53(3):569--576, March 1966.

\bibitem{Steward1991In}
K.~L. Steward and T.~Linn.
\newblock {In vivo analysis of overlapping transcription units in the
  rplKAJLrpoBC ribosomal protein-RNA polymerase gene cluster of Escherichia
  coli.}
\newblock {\em Journal of molecular biology}, 218(1):23--31, March 1991.

\bibitem{Szklarczyk2014STRING}
Damian Szklarczyk, Andrea Franceschini, Stefan Wyder, Kristoffer Forslund,
  Davide Heller, Jaime Huerte-Cepas, Milan Simonovic, Alexander Roth, Alberto
  Santos, Kalliopi~P. Tsafou, Michael Kuhn, Peer Bork, Lars~J. Jensen, and
  Christian von Mering.
\newblock {STRING v10: protein-protein interaction networks, integrated over
  the tree of life.}
\newblock {\em Nucleic acids research}, October 2014.

\bibitem{Tatusov1997Genomic}
R.~L. Tatusov, E.~V. Koonin, and D.~J. Lipman.
\newblock {A genomic perspective on protein families.}
\newblock {\em Science (New York, N.Y.)}, 278(5338):631--637, October 1997.

\bibitem{Ward2014Quickly}
Natalie Ward and Gabriel Moreno-Hagelsieb.
\newblock {Quickly Finding Orthologs as Reciprocal Best Hits with BLAT, LAST,
  and UBLAST: How Much Do We Miss?}
\newblock {\em PLoS ONE}, 9(7):e101850+, July 2014.

\bibitem{Wolf2001Genome}
Y.~I. Wolf, I.~B. Rogozin, A.~S. Kondrashov, and E.~V. Koonin.
\newblock {Genome alignment, evolution of prokaryotic genome organization, and
  prediction of gene function using genomic context.}
\newblock {\em Genome research}, 11(3):356--372, March 2001.

\end{thebibliography}
\end{document}